# A POSSIBLE OPTICAL COUNTERPART FOR THE TRIPLE PULSAR SYSTEM PSR B1620−26

Charles D. Bailyn[1,2], Eric P. Rubenstein[1,3], Terrence M. Girard[1], Dana Dinescu[1], Frederic A. Rasio[4,5], Brian Yanny[4].

ABSTRACT

We have searched deep optical images of the globular cluster M4 for a possible optical counterpart for the third body in the PSR B1620−26 system. We identify as a possible optical counterpart a star located within $0.3''$ of the nominal pulsar position with $V \approx 20$. This magnitude is consistent with a main sequence cluster member with $M \approx 0.45 M_\odot$. However, this star may be blended with fainter stars, and the probability of a chance superposition is non-negligible.

1. Dept. of Astronomy, Yale University, PO Box 208101, New Haven, CT 06520-8101.
2. National Young Investigator
3. Visiting Astronomer, Cerro-Tololo Interamerican Observatory. CTIO is operated by the Association of Universities for Research in Astronomy (AURA) Inc., under contract with the National Science Foundation.
4. Institute for Advanced Study, Olden Lane, Princeton NJ, 08540.
5. Hubble Fellow.

## I. Introduction

M4 is the nearest globular cluster to Earth, and one of the first in which a millisecond pulsar was discovered (Lyne et al. 1988). The pulsar, PSR B1620−26, is in a nearly circular binary system with a low mass companion of $\approx 0.3 M_\odot$. Millisecond pulsars in such configurations are not uncommon in and out of globular clusters, although the orbital eccentricity of PSR B1620−26 ($e = 0.0253$) is several orders of magnitude larger than in most low mass binary pulsars (McKenna & Lyne 1988). Optical emission from such systems is expected to be small, as is the case for PSR J0437−4715, which has a distance of $\approx 150$ parsecs and a visual magnitude of $V \approx 20.7$ (Bailyn 1993; Bell, Bailes & Bessell 1993). Such an object would be undetectable in any globular cluster.

Recently, however, the timing data for PSR 1620−26 have revealed a very large value of the pulse period second derivative, indicating the presence of an additional source of acceleration (Backer, Foster & Sallmen 1993; Thorsett, Arzoumanian, & Taylor 1993). The most likely explanation of this anomalous acceleration is that the pulsar has a second more distant orbital companion. Such hierarchical triple systems are expected to be produced quite easily in globular clusters through dynamical interactions (Mikkola 1984; Bailyn 1989; Hut 1992). Current timing data allow the third object to be of either planetary or stellar mass (Backer et al. 1993; Thorsett et al. 1993). In the latter case, the third object might be optically detectable. Given the excellent astrometry available from timing of pulsars, we thought it worthwhile to search deep images of the cluster for an optical counterpart.

## II. Astrometry

Due to their small 6'x6' field of view, the CCD images discussed below contain no astrometric standards. Therefore we used two photographic plates taken in May and June 1972 with the Yale Southern Observatory 51-cm double astrograph at El Leoncito, Argentina to provide a link between the CCD and celestial coordinates. The plates have a scale of $55.1''/\text{mm}$ and cover a region $3.8 \times 3.1$ degrees, centered on M4. The four-minute exposure of each plate reaches approximately 15th magnitude.

Approximately 140 stars on the CCD frames were selected whose images were deemed well-isolated and measurable on the plates. These stars would link the CCD coordinates to the plate coordinates. We also measured 90 stars from the PPM Catalog (Roeser & Bastian, 1993), 11 of which are also in the IRS Catalog (Corbin, 1991). These stars would function as the primary reference stars, defining the transformation from plate measures to celestial coordinates.

The two YSO plates were measured with the Yale PDS microdensitometer in fine-scan mode. In addition to the program stars, five stars were repeatedly measured throughout the scan to correct for slight drifts in the PDS metric system as a function of time. Image centers and pseudomagnitudes were calculated using our standard 2-D Gaussian fitting routine, (Lee and van Altena, 1983). An additional culling of bad plate images was performed by visual inspection of the digitized images. Five of the eleven IRS stars had to be eliminated due to saturated or blended images.

Photographic positions often display a magnitude-dependent bias in the image positions due to telescope guiding error. There is also evidence for lens decentering in the YSO astrograph, which gives rise to a similar effect. To test for the presence of a magnitude bias in our measures, we performed a linear transformation of the plate coordinates into those derived by ALLSTAR for the CCD frame and plotted coordinate residuals as a function of magnitude. The linear response of the CCD should make it free of magnitude bias, short of saturation effects. The residual plots showed a hint of structure, particularly in the x-coordinate at intermediate magnitudes, but at a level that could not be confidently modelled. Therefore we have made no explicit correction of the plate measures for magnitude.

The CCD star coordinates were transformed to the plate coordinate system, using separate reductions for each of the two plates. The linear transformation, using $\approx 120$ stars, had a single-coordinate, unit-weight standard error of $0.12''$, for each plate. Thus, the formal error of the mean transformation is $\approx 0.01''$. Next, the transformation from plate coordinates to celestial coordinates was derived using the primary reference stars. Catalog proper motions were used to bring the PPM and IRS star positions to the plate epoch and then transformed to standard coordinates by gnomonic projection. Using $\approx 70$ PPM stars to define the reference frame, a transformation involving linear and quadratic plate-tilt terms was found by least-squares solution. The single-coordinate standard error of the solution was $\approx 0.12''$, yielding an uncertainty in the mean transformation of $\approx 0.014''$. A trial solution, using only linear terms, was also made with the six measurable IRS reference stars, whose positions should be of somewhat higher accuracy than those of the PPM. The differences between the solutions based on the two different catalogs were insignificant and the PPM solution was adopted as final.

The coordinates of the pulsar are $\alpha_{2000}=$ 16h 23m 38.223s, $\delta_{2000}=$ -26d 31' 53.71" at approximate epoch 1990, (Thorsett et al. 1993). We use the Cudworth and Rees (1990) value of the proper motion of the cluster to determine a position for the pulsar at the epoch of the YSO plates. The CCD coordinates for the pulsar calculated using the transformations separately derived for the two plates agree to within $0.03''$. This agreement supports the conclusion that these plates do not suffer from any significant magnitude equation.

While the formal random error introduced by the plate transformations detailed above is small, the true error will actually be dominated by systematic effects. By comparison to the HIPPARCOS preliminary (30 month) catalog (I. Platais, priv. comm.), we estimate the maximum possible error from a magnitude dependent effect due to lens decentering to be $0.1'' = 0.25$ pixels, over the four-magnitude difference in mean V between the PPM stars and the stars used to link the CCD and plate material. Another possible source of error is in the adopted absolute proper motion of M4, a very difficult quantity to measure. Since the relevant epoch difference is only 20 years, we estimate the maximum contribution from this uncertainty to be $< 0.05'' = 0.12$ pixels, based on a pessimistic tripling of the error estimate of Cudworth and Rees (1990). A final uncertainty resides in the underlying assumption that the adopted right ascension and declination of the pulsar, determined from radio timing measurements, can be compared directly to the optical reference frame of the PPM. The solar system ephemeris on which the radio positions are based is thought to be accurate to $0.05''$ (Standish 1990), while the PPM catalog may have systematic errors of similar magnitude (Roeser 1990). Summing all of these possible effects, we find a maximum systematic error in determining the pulsar position on the CCD of $\approx 0.25''$.

### III. Photometry

On June 20-22 and 29/30 1993, a large number of CCD images of the core of M4 were obtained at the Cerro-Tololo Interamerican Observatory 0.9m telescope, equipped with a Tek1024 CCD. There were 39 images obtained in B, with typical exposure times of 300s, and 75 images obtained in V, with typical exposure times of 150s. The advantage of using an average of many short exposures rather than a single long exposure is that the brighter stars are not saturated, and can therefore be subtracted digitally to reveal the fainter stars beneath them. The exposure times were chosen to barely saturate the brightest giants in the cluster. Conditions were non-photometric, so some exposures were lengthened due to patches of relatively high extinction. The individual images were flatfielded and debiased using standard IRAF routines. Cosmic rays were removed from the images using FIGARO routines, and then the images were registered and averaged. Seeing in the averaged images was $\approx 1.5$ arcseconds FWHM. The region near the nominal pulsar position is shown in Figure 1.

We then performed an analysis of the entire image using Peter Stetson's programs DAOPHOT2 and ALLSTAR2 (Stetson, Davis & Crabtree 1991). The point spread function (PSF) was in the form of a modified Penny function, the parameters of which were allowed to vary cubically across the field, and a table of residuals. Over 60 stars in each image were used to define the PSF, and nearly a thousand near neighbor stars were subtracted from the image to produce clean profiles for these stars. After iterating a sequence of neighbor star subtractions and PSF determinations, we arrived at a final PSF. When this PSF was used to subtract the original PSF stars and their neighbors from the images, the largest residual was 0.5% of the peak intensity of the PSF star in question, and typical residuals were $\approx 0.1\%$. Over $2 \times 10^4$ stars in common between the two colors were identified. Outside of the central 1' of the cluster, the principle stellar sequences were determined down to 3.5 magnitudes below the main sequence turnoff. A detailed analysis of the full color magnitude diagram will be reported elsewhere. Since the observing conditions were not photometric, we have put our instrumental magnitude scale onto a standard scale by comparing our color-magnitude diagram with previous work (Cudworth & Rees 1990), with resulting zero-point errors of about 0.1 magnitude.

We analyzed the region within 6" of the pulsar position with particular care. The residuals from our preliminary photometry were examined by eye, and additional stellar objects identified. Then the profile fitting routine was rerun, and the process iterated until no further stellar objects in this region could be identified. Results of this process are shown in Figure 2, and the positions, magnitudes, and colors of the stars detected are listed in Table 1.

A star (labelled star 1 in Table 1) was identified $\approx 0.3''$ south of the nominal pulsar position. As can be seen in Figure 1, the star's image is not symmetric, an effect which could be due to background fluctuations, residuals in the fits to nearby brighter stars, or blending with fainter stars. Any of these effects will cause errors in determining the star's centroid which could be as large as half a pixel, or $0.2''$. Given the errors in the astrometry described above, star 1 could be associated with PSR 1620−26. However, given the density of stars found in this part of the cluster, the chance probability of a star being found within $0.3''$ of *any* arbitrary position is non-negligible. To quantify the probability of a chance superposition, we performed a Monte-Carlo simulation in which we compared random positions within 6" of the nominal pulsar position with the list of stars given in Table 1. We found that 320 out of $10^4$ chance positions were within 0.3" of one of the stars on the list. Thus the possibility of a chance superposition is $\approx 1/30$, and should not be discounted. A star with the V magnitude of star 1 (the B color is highly uncertain) in a cluster with the distance, reddening and metallicity of M4 is compatible with being a cluster main sequence star with $M \approx 0.45 M_\odot$ (Green, Demarque & King 1987).

### IV. Discussion

The two largest sources of uncertainty in the position of the pulsar in our optical images are the link betwen the radio and optical coordinate systems, and the absolute proper motion of the cluster. The former difficulty will be significantly reduced when HIPPARCOS data become available. The latter problem could be eliminated by using current epoch plate material. Higher resolution CCD images would decrease the centroiding error for star 1, and determine whether it is a single star or a blend, (as is suggested by the assymetric shape of the image) and if there are other fainter candidate stars.

If star 1 is indeed the second companion of PSR B1620−26, its estimated mass of $\approx 0.45 M_\odot$ would imply an outer orbit with a fairly large eccentricity ($e_2 \geq 0.5$ to be consistent with the radio timing data — see Michel 1994). Such a large eccentricity agrees nicely with the predictions of dynamical calculations of triple formation through interactions between two binary systems (Rasio, Hut & McMillan 1994). In addition, the presence of a second companion of stellar mass (as opposed to planetary mass) would naturally explain the anomalous eccentricity of the (inner) binary pulsar as arising from secular orbital perturbations (Mazeh & Shaham 1979, Rasio 1994).

We are grateful for conversations with W. van Altena, Z. Arzoumanian, and I. Platais, and for assistance in performing the CCD observations from W. Sherry and members of the CTIO staff. We are grateful to D. Maoz of the Wyse Observatory for sharing additional images of the cluster. CDB and EPR acknowledge support from NASA grant NAGW-2469. TMG and DD acknowledge support from grants from NASA and the NSF. FAR and BY are supported by Hubble Fellowships, funded by NASA through grant HF-1037.01.92A and HF-1013.01-90A respectively, from STScI, which is operated by AURA under contract NAS5-26555.

TABLE 1

| id # | " N of pulsar | " E of pulsar | V | B − V |
|---|---|---|---|---|
| 1 | -0.32 | 0.03 | 20.04 | * |
| 2 | 1.57 | 0.24 | 18.95 | * |
| 3 | 1.69 | -0.69 | 16.16 | 1.03 |
| 4 | -0.57 | 2.78 | 18.28 | 1.08 |
| 5 | -2.78 | -0.70 | 19.92 | * |
| 6 | -0.10 | -2.96 | 17.93 | 1.09 |
| 7 | 1.75 | 2.77 | 17.28 | 0.95 |
| 8 | -1.15 | 3.72 | 18.53 | * |
| 9 | -0.61 | 4.49 | 17.81 | 1.01 |
| 10 | 3.91 | -2.58 | 16.07 | 0.91 |
| 11 | -4.64 | -0.92 | 16.90 | 0.88 |
| 12 | 5.01 | 0.17 | 13.90 | 1.20 |
| 13 | -5.48 | 0.90 | 17.26 | 0.90 |
| 14 | -4.71 | 3.42 | 16.86 | 0.90 |
| 15 | -5.27 | -2.87 | 16.49 | 0.88 |
| 16 | 5.62 | 3.25 | 17.55 | 1.05 |

* Although they were detected in the B frame, accurate colors could not be determined for these stars.

**Figure Captions**

Figure 1: V-band contour plot of the 60 × 60 pixel ($\approx 24'' \times 24''$) region centered on the pulsar position. East is up and north to the right. The three brightest stars listed in Table 1 are indicated.

Figure 2: Top panel shows the V image of the same region as Figure 1 with the stars listed in Table 1 digitally subtracted. Second panel represents the same data as the second, but with equally spaced linear contours covering the range from the mean background level near the pulsar to the second contour in the second panel. Note that *all* the stars visible in Figure 1 are above the maximum contour level of this plot. Bottom panel has the same contouring as the third panel, but with star 1 unsubtracted.